# The mass of cosmic rays above $10^{17}$ eV

M. T. Dova[a], A.G. Mariazzi[b], A. A.Watson[b]

*(a) Departamento de Fisica, Universidad Nacional de La Plata,CC67 La Plata (1900), Argentina*
*(b) School of Physics and Astronomy, University of Leeds, Leeds LS2 9JT, UK*
Presenter: A. G. Mariazzi (agm@ast.leeds.ac.uk), uki-analisa-M-abs1-he14-oral

Interpretation of the energy spectrum and arrival distribution of cosmic rays is complicated by lack of knowledge of the nature of the primaries. We review claims for the mass composition above $10^{17}$ eV where it can be determined only indirectly from air-shower observables. Difficulties in comparisons between data arise because, inevitably, a set of measurements is interpreted using the best model of hadronic interactions available at the time of analysis. We discuss the situation and conclude that the evidence for a proton-dominated mass composition, even at the highest energies, is unconvincing. However, it may be that there are consistent differences between mass measurements from optical techniques and those based upon other shower observables. We also find that iron nuclei of ultra high energy can probably escape from the galaxies that host GRBs, possible cosmic ray accelerators. The accelerators must lie nearby.

## 1. Introduction

Our knowledge of the mass of primary cosmic rays at energies above $10^{17}$ eV is rudimentary. Reliable information about the mass composition is essential if we are to interpret data on the energy spectrum and arrival directions that is accumulating from several experiments. Above $\sim 10^{15}$ eV, virtually all information derives from studies of the properties of air showers and inferences rely, in all cases, on interpretations based on model calculations. As data from particular experiments are rarely analysed contemporaneously, some of the differences reported arise from differences between the models that are in vogue at a particular time. Some of the points highlighted in this report have already been made in [1]: others are new.

Methods of determining the composition of the primary can be classified into 3 groups: (i) measurements of $X_{max}$ and fluctuations in the distribution of $X_{max}$ made from air-Cherenkov light or fluorescence light emissions, (ii) muon densities and (iii) measurements that are essentially geometrically-based, such as slope of the lateral distribution function (LDF) or the thickness of the shower-front or the shower front curvature.

## 2. A brief overview of the experimental technique

The Fly's Eye group pioneered the direct measurement of shower maxima ($X_{max}$). New results have been reported by the HiRes group [2] who have determined $X_{max}$ from $10^{18}$ to $10^{19.4}$ eV. In addition, they have measured the fluctuation of $X_{max}$ as a function of energy. Using the QGSJET01 model they interpret their results as being consistent with 'a constant or slowly changing, predominantly light composition'. The Yakutsk group [3] has made similar measurements using air-Cherenkov light. The deduction of the depth of maximum is less direct in this case but the conclusions are very similar. For measurements that use optical techniques, careful monitoring of the atmosphere is essential, the more so in the case of fluorescence light where the emission is isotropic and details of the temperature and pressure at the emission point are crucial because of electron recombination times. Also the magnitude of the fluorescence yield remains an issue.

It is well-known that a shower produced by an iron nucleus will contain a greater fraction of muons at the observation level than a shower of the same energy created by a proton primary. Many efforts to derive the mass spectrum of cosmic rays have been attempted over the full range of air shower energies using this fact.

However, although the differences are predicted to be relatively large (on average there are $\sim 70\%$ more muons in an iron event than a proton event), there are large fluctuations and, again, there are differences between what is predicted by different models. Thus, the QGSJET model set predicts more muons than the SIBYLL family, the difference arising from different predictions as to the pion multiplicities produced in nucleon-nucleus and pion-nucleus collisions that in turn arise from differences in the assumptions about the parton distribution within the nucleon [4]. In figure 1, data from the Yakutsk and AGASA groups are compared above $10^{19}$ eV. To make this comparison we have adjusted the Yakutsk data ($E_\mu > 1$ GeV, [3]) to that of AGASA ($E_\mu > 0.5$ GeV, [5]) using the relations given in [4]. No correction has been made for the difference in altitude (1020 to 920 g cm$^{-2}$) but the agreement is reasonably good. However, particularly at the very highest energies, it is clear that neither data set is able to discriminate between p or Fe as the primary.

We derived an upper limit for iron primaries above $10^{19}$ eV of 64 % using Yakustk data following the method described in [6]. For each event a chisquare value was derived by comparing the observed muon densities with those predicted for iron primaries from simulations with the hadronic interaction model QJSJET01[3].

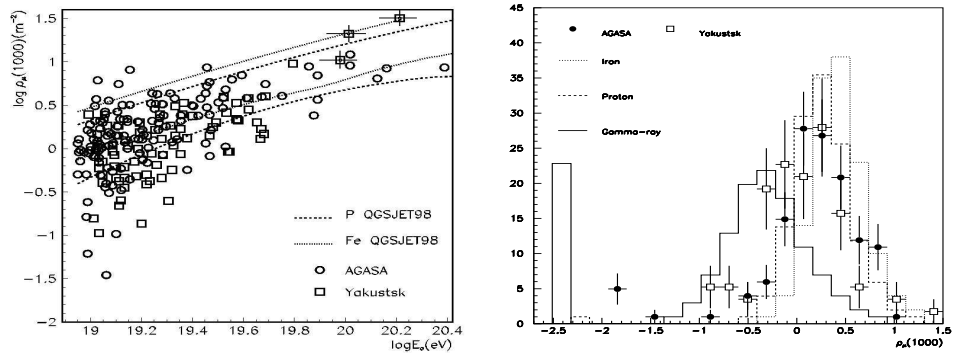

**Figure 1.** The left plot is the muon density at 1000 m vs. E for AGASA events (circles) and for Yakutsk events (squares). Expected 1 $\sigma$ bounds for proton, iron and photon primaries (from QGSJET98) are indicated. The corresponding muon density distributions, summed over energy, are on the right.

Measurements of the lateral distribution function (LDF), a geometric parameter, were made with high precision at Volcano Ranch [7] and Haverah Park [8]. Both the average value, but more usefully, the fluctuations in the LDF can be compared with model calculations to derive a primary mass. The fraction of Fe nuclei reported in [8] is 52 % while in [1] a larger fraction ($\sim 75$ %) is inferred but over an energy range that is uncertain.

Another 'geometrical' quantity that has been used to infer the mass composition is the thickness of the shower disc as characterised by $t_{1/2}$, the 10-50 % risetime. Using Haverah Park data a mass composition that is rather heavy has been deduced, using the QGSJET01 model, for a sample of events of energy $10^{19}$ eV [9].

## 3. Summary and discussion

We have attempted to summarise the information reported in the work discussed above in figure 2. This diagram is an extension of one shown in [1]. We have added the deductions of the Yakutsk group made from their measurements of muons and air-Cherenkov light and the apparently conflicting data from the HiRes/MIA experiment [10]. There is clearly an immense spread in the estimates of the Fe fraction. As noted in [1] (and independently by Engel and Klages [12]) there is a systematic difference in the inferences drawn from the methods that can assess $X_{max}$ directly and from those that use muon densities or geometrically-based methods. It is not clear whether this is real or if it is an artifact of the measurements or the analyses. We note also that the next generation of models (QGSJETII [13]) is expected to predict a smaller number of muons for a particular mass. Engel [14] has indicated that the predicted muon density between 600 and 1000 m might fall by 30 %. It is also predicted [14] that the depth of maximum will be deeper in the atmosphere by 10 and 20 g cm$^{-2}$ for proton and iron-initiated showers respectively. While these changes are insufficient to reconcile the optical data with that from muons, risetimes and LDFs, they are in the sense as to imply a heavier mass composition at high energies. In particular claims for a dominantly protonic mass composition seem likely to be weakened. Furthermore the narrowing of the difference between the $X_{max}$ predictions makes discovering the mass by this method even harder.

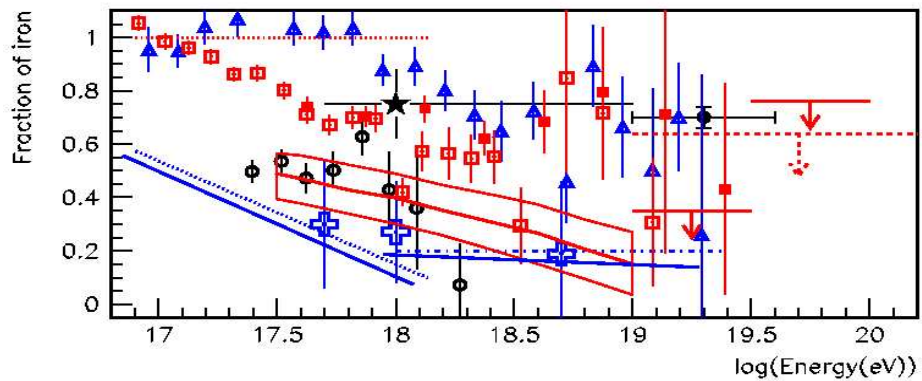

**Figure 2.** Iron fraction from various experiments. Muon densities (red): AGASA A1 (solid line with band) and upper limits above $10^{19}$ eV (solid lines) with QGSJET98 [5]; AGASA A1 (squares) and A100 (filled squares) deduced using Mocca-SIBYLL [11]; Yakutsk upper limit (dashed line) using QGSJET01 [3]. MIA (dotted line) QGSJET98 iron fraction estimate. Geometrically-based methods (black): Volcano Ranch LDF (star) [1], Haverah Park LDF (circles) [8] and risetime (point) [9] with QGSJET01. $X_{max}$ (blue): HiRes (dot dash line) [2] and Yakutsk (crosses) [3] derived from fluctuation in the $X_{max}$ distributions. HiRes (triangles) using Mocca-SIBYLL [11], HiRes-MIA (dotted line) [10] using QGSJET98, HiRes (solid lines) using QGSJET01 [2] deduced from $X_{max}$ mean values.

## 4. Can heavy nuclei escape from the production site?

We believe that the evidence summarized in Figure 2 shows that the common assumption that protons dominate the mass spectrum at the highest energies may not be correct. A mixed-composition, containing a mixture of species that will reflect what was accelerated and the propagation from the sources to the earth, seems likely. Heavy nuclei are interesting candidates as UHECR primaries as this would help to reconcile the AGASA/HiRes differences of energy scale [15] . Also they are easier to accelerate and the greater deflections in magnetic fields could explain the isotropy of the arrival directions.

A currently-popular idea for an accelerator is a gamma ray burst (GRB)[16]. It is possible that the GRB shocks might accelerate galactic material surrounding the source that contains heavy nuclei [17]. An interesting question therefore arises: is it possible for a heavy nucleus to escape photo-disintegration as it exits the galaxy? We assume that the source can accelerate Fe nuclei to $10^{21}$ or $10^{22}$ eV. The photo-disintegration process is dominated by the giant dipole resonance, which peaks at $\sim$ 20 MeV for Fe. Thus mm-wavelength photons of $\sim 2.5\,10^3$ and $\sim 2.5\,10^4$ $\mu$m are relevant for photo-disintegration. A typical photo-disintegration cross-section is 40 mb. Recent studies of the properties of the host galaxies of GRB reveal indirect evidence for the link between GRBs and star formation. The host of the GRB 980703 is at the faint end of the class of ultra-luminous infrared galaxies, with $L_{FIR} \sim 10^{12}$ $L_\odot$ near the center of the galaxy in a region of star formation[18].

The luminosity per unit of frequency for these radio wavelengths is approximately $L_\nu=10^{30}$ ergs$^{-1}$Hz$^{-1}$ and $L_\nu=3\,10^{30}$ ergs$^{-1}$Hz$^{-1}$ [18]. Radio and mm observations show that the host galaxy luminosity arises in an innermost disk of r$\sim$0.2 kpc to 2kpc [19]. The typical photon densities at $2.5\,10^3$ and $\sim 2.5\,10^4$ $\mu$m are thus $10^3$ and $3.10^3$ cm$^{-3}$ and the corresponding mean free paths for photo-disintegration are 8 kpc and 2.7 kpc respectively. It seems possible, therefore, that heavy nuclei accelerated by a GRB to energies of $10^{21}$ eV or $10^{22}$ eV, could escape from the host galaxy of the GRB. Of course, iron nuclei can be fragmented as they travel to earth. Complete fragmentation is predicted to occur within 10 Mpc [20] so that, if the primary beam does contain a mixture of nuclei, sources, even at the highest energies, must be nearby.

## 5. Conclusions

The highest energy cosmic rays may be nuclei rather than protons. Fe-nuclei, if accelerated in the galaxies that host GRBs, should be able to leave the galaxy before photo-disintegration. The accelerators must be nearby.

## 6. Acknowledgements

We thank Stuart Lumsden and Tom Hartquist for illuminating discussions about the host galaxies of GRBs.